\documentclass[conference, twocolumn, 9pt]{IEEEtran}
\usepackage{fixltx2e-fixes}
\usepackage{IEEEtran-fixes}
\usepackage[utf8x]{inputenc}
\usepackage[T1]{fontenc}
\usepackage{textcomp}
\usepackage[T1]{url}
\usepackage[final]{microtype}
\usepackage{amsmath}
\usepackage[thmmarks, amsmath]{ntheorem}
\usepackage[romanConstants, boldVectors]{scstuff}
\usepackage{units}
\usepackage{xspace}
\usepackage{bm}
\usepackage{mdwtab}
\usepackage{acronym}
\usepackage{cite}
\usepackage[pdftex]{graphicx}
\usepackage[svgnames]{xcolor}
\usepackage[italian,english]{babel}
\usepackage[caption=false]{subfig}
\usepackage[unicode,pdftex]{hyperref}
\usepackage{paralist}

\hypersetup{colorlinks=true, 
  linkcolor=DarkBlue,
  citecolor=DarkBlue,
  urlcolor=DarkBlue}

\graphicspath{{./}{../Figures/}{./Figures/}}

\newcommand{\emailurl}[1]{%
  \href{mailto:#1}{\nolinkurl{#1}}}

\newacro{NDT}{Non Destructive Testing}
\newacro{MIMO}{Multiple Input Multiple Output}
\newacro{MUT}{Material Under Test}
\newacro{eSNR}{echo SNR}
\newacro{PDS}{Power Density Spectrum}
\newacro{CDMA}{Code Division Multiple Access}
\newacro{PN}{Pseudo Noise}
\newacro{FM}{Frequency Modulation}
\newacro{PAM}{Pulse Amplitude Modulated}
\newacro{PDF}{Probability Density Function}

\newcommand{\Df}{\ensuremath{\Delta f}}
\newcommand{\pamperiod}{\ensuremath T_{c}}
\newcommand{\Tc}\pamperiod
\newcommand{\spectrum}[1]{\ensuremath{\Phi_{#1#1}}}
\newcommand{\TB}{\ensuremath{\mathit{TB}}}

\theorembodyfont{\itshape}
\theoremheaderfont{\normalfont\bfseries}
\theoremseparator{}

\theoremstyle{nonumberplain}
\theoremheaderfont{\normalfont\bfseries}
\theorembodyfont{\normalfont}
\theoremsymbol{~~\IEEEQEDopen}

\makeatletter
\let\thanks\@IEEESAVECMDthanks
\makeatother

\PrerenderUnicode{©}

\renewcommand\normalsize{\fontsize{9.25}{11.12}\selectfont}

\begin{document}

\title{%
  From Chirps to Random-FM Excitations in Pulse Compression Ultrasound Systems}
                              
%
\author{%
  \IEEEauthorblockN{%
    Sergio Callegari\IEEEauthorrefmark{1},
    Marco Ricci\IEEEauthorrefmark{2},
    Salvatore Caporale\IEEEauthorrefmark{1},
    Marcello Monticelli\IEEEauthorrefmark{1},\\
    Massimiliano Eroli\IEEEauthorrefmark{2},
    Luca Senni\IEEEauthorrefmark{2},
    Riccardo Rovatti\IEEEauthorrefmark{1},
    Gianluca Setti\IEEEauthorrefmark{3} and
    Pietro Burrascano\IEEEauthorrefmark{2}}\vspace{1ex}
  \IEEEauthorblockA{%
    \IEEEauthorrefmark{1}ARCES/DEIS, University of Bologna, Italy\\
    Email: \emailurl{sergio.callegari@unibo.it},
    \emailurl{salvatore.caporale@unibo.it},\\
    \emailurl{marcello.monticelli@studio.unibo.it},
    \emailurl{riccardo.rovatti@unibo.it}}\vspace{0.8ex}
  \IEEEauthorblockA{%
    \IEEEauthorrefmark{2}Polo Scientifico Didattico di Terni, 
    University of Perugia, Italy\\
    Email: \emailurl{marco.ricci@unipg.it},
    \emailurl{massimiliano.eroli@unipg.it},
    \emailurl{luca.senni@unipg.it},
    \emailurl{pietro.burrascano@unipg.it}}\vspace{0.8ex}
  \IEEEauthorblockA{%
    \IEEEauthorrefmark{3}ENDIF, University of Ferrara, Italy\\
    Email: \emailurl{gianluca.setti@unife.it}}
  \thanks{This is a post-print version of a paper published in the Proceedings
    of the 2012 IEEE International Ultrasonics Symposium (IUS), 2012, pp.\@ 471
    - 474, Dresden (DE), Oct.\@ 2012. Available through DOI
    \href{http://dx.doi.org/10.1109/ULTSYM.2012.0117}%
    {10.1109/ULTSYM.2012.0117}. To cite this document, please use the published
    version data.}
  \thanks{Copyright © 2012 IEEE. Personal use of this material is
    permitted. However, permission to use this material for any other purposes
    must be obtained from the IEEE by sending a request to
    \url{pubs-permissions@ieee.org}.\protect\\[-2ex]}}

\initializeplaintitle[%
  \def\texorpdfstring#1#2{#2}]
\hypersetup{%
  pdftitle=\plaintitle),
  pdfauthor={Sergio Callegari, Marco Ricci, Salvatore Caporale, Marcello
    Monticelli, Massimiliano Eroli, Luca Senni, Riccardo Rovatti, Gianluca
    Setti, Pietro Burrascano}
}

\maketitle

\begin{abstract}
  Pulse compression is often practiced in ultrasound \ac{NDT} systems using
  chirps. However, chirps are inadequate for setups where multiple probes need
  to operate concurrently in \ac{MIMO} arrangements. Conversely, many coded
  excitation systems designed for \ac{MIMO} miss some chirp advantages
  (constant envelope excitation, easiness of bandwidth control, etc.) and may
  not be easily implemented on hardware originally conceived for chirp
  excitations. Here, we propose a system based on random-FM excitations, capable
  of enabling \ac{MIMO} with minimal changes with respect to a chirp-based
  setup. Following recent results, we show that random-FM excitations retain
  many advantages of chirps and provide the ability to frequency-shape the
  excitations matching the transducers features.
\end{abstract}
\acresetall
\acused{SNR}
\acused{rms}
\acused{ac}
\acused{A/D}
\acused{D/A}
\acused{FM}

\section{Introduction}
\label{sec:intro}
Many ultrasonic frameworks for \ac{NDT} rely on pulse compression to cope with
highly dissipative materials or with setups where large dissipation is
encountered at the probe interfaces \cite{Gan:Ultrasonics-39-3}. The approach
extends the excitation time-bandwidth product to enhance the \ac{eSNR} without
sacrificing resolution at the cost of some signal processing typically
practiced by \emph{matched filters} \cite{Turin:IRETIT-6-3}.

In order to maximize the \ac{eSNR} one should feed the \ac{MUT} with a large
acoustic energy. At the same time, overloading the probes or exceeding acoustic
pressure limits, even transiently, could result in distortion. These
requirements make constant envelope excitations an attractive
feature. Consequently, \emph{chirps} are a widely adopted large-bandwidth
excitation. Chirps also have the advantage of easing bandwidth control. For a
linear, slow chirp, the \ac{PDS} is almost uniform within the two extreme
frequencies and almost null outside. This makes it straightforward to carefully
match the probes bandwidth.

Nonetheless, chirps cannot be considered a universal solution. In some setups,
it is desirable to rely on multiple transmitting probes that are simultaneously
operated (e.g., to reduce test time, or to deal with situations where the
\ac{MUT} moves as in some automated testing facilities.). Wide-band excitations
such as those needed for pulse compression offer a natural opportunity to
create families of excitations where the members are highly orthogonal and thus
suited for \ac{MIMO} operation. This \emph{code-division} approach sets a
parallel with some tele-communication schemes
\cite{Callegari:CAT-2005-5-alt}. Yet, chirps are clearly not immediately suited
for this approach.

Thus, different types of excitations have been developed.  \emph{Coded
  excitation} are often based on \ac{PN} sequences ($m$-sequences, Kasami, Gold
code families, etc.) which have proved quite fruitful since the seminal work of
Golomb \cite{Golomb:SDGC-2005}.  Discrete-valued (often binary) sequences are
either used directly or upconverted/modulated into ultrasound excitations
\cite{Burrascano:CSIE-2009}. Good \emph{$\delta$-like} auto-correlation
properties guarantee good time resolution. Extension in time assures that a
sufficient amount of energy is pushed into the \ac{MUT}. Finally, low
cross-correlation assures reduced mutual interference.  In spite of these
advantages, pitfalls exist. For instance, modulated sequences often have a
non-constant envelope. Thus, to fit inside the maximum power range of the
probes, one ends up with an average power lower than the maximum one, wasting
power conversion ability. Secondly, difficulties may emerge in bandwidth
control and in containing energy inside the probe bandwidth. Finally, this
excitations may require a rather different hardware than chirps, where an
\ac{FM} block fed by a ramp is often all that is needed at the transmitters.

Here, we propose a solution that may be seen as a bridge between chirp-based
excitations and coded excitations. It is based on the \ac{FM} modulation of
\ac{PAM} sequences so that hardware changes with respect to chirp based systems
can be minimal and constant envelope operation can be guaranteed. It offers
sufficiently good cross-correlation properties for \ac{MIMO}. It can be
extended to differentiate excitations based on discrete codes. It provides an
easy adjustment to the probe bandwidth. Furthermore, it has advantages of its
own, like the possibility to carefully shape the \ac{PDS}, following
\cite{Callegari:TCAS1-50-1}. Similarly to nonlinear chirps
\cite{Ricci:IUS-2012} this lets one equalize the probe response, to enhance the
actual bandwidth (and thus resolution) or transferred power (and thus
\ac{eSNR}) \cite{Pollakowski:TUFFC-41-5}.

\section{Random-FM excitations}
\subsection{Background on chirps and random-FM signals}
\label{ssec:background}
A chirp is defined by the following modulation
\begin{equation}
  s(t) = \Re\left(e^{2 \pi i(f_0 t + \Df \int_{-\infty}^{t} x(\tau)\,
      d\tau)}\right)
  \label{eq:FM}
\end{equation}
where $A$ is the envelope amplitude, $f_0$ is the central frequency, $\Df$ is
the maximum frequency deviation from $f_0$, and $x(t)$ is a monotonically
increasing, smooth modulating signal taking values in $[-1,1]$.  The simplest
case is evidently that of a linear chirp in which $x(t)=-1+\nicefrac{2 t}{T_e}$
where $T_e$ is the chirp length.  As long as $T_e$ is sufficiently large with
respect to $\nicefrac{1}{\Df}$ (i.e., the modulation is \emph{slow} enough),
the energy of the linear chirp is almost uniformly distributed in $[f_0-\Df,
f_0+\Df]$ and almost null elsewhere, which makes it an effective excitation.

\begin{figure}
  \begin{center}
    \subfloat[\label{sfig:chirp}]{%
      \shortstack{%
        \includegraphics[scale=0.55]{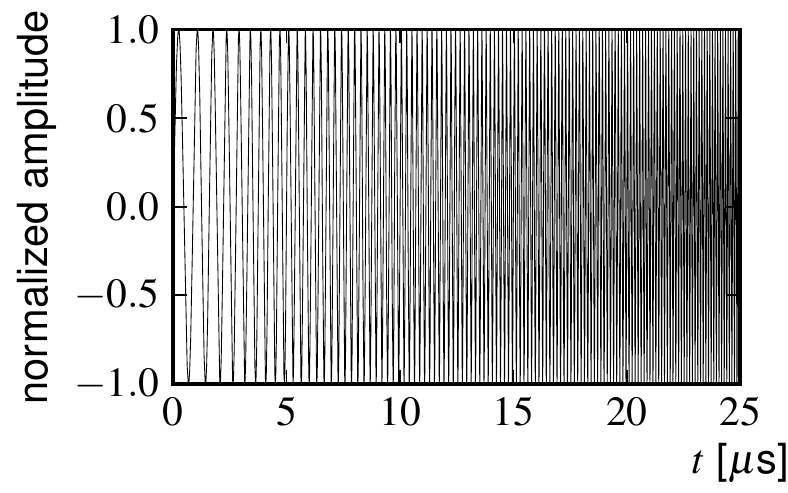}\\
        \includegraphics[scale=0.55]{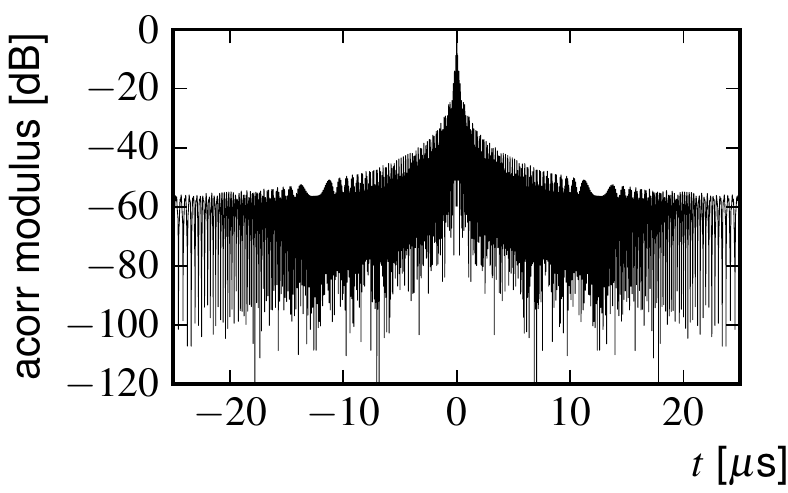}\\
        \includegraphics[scale=0.55]{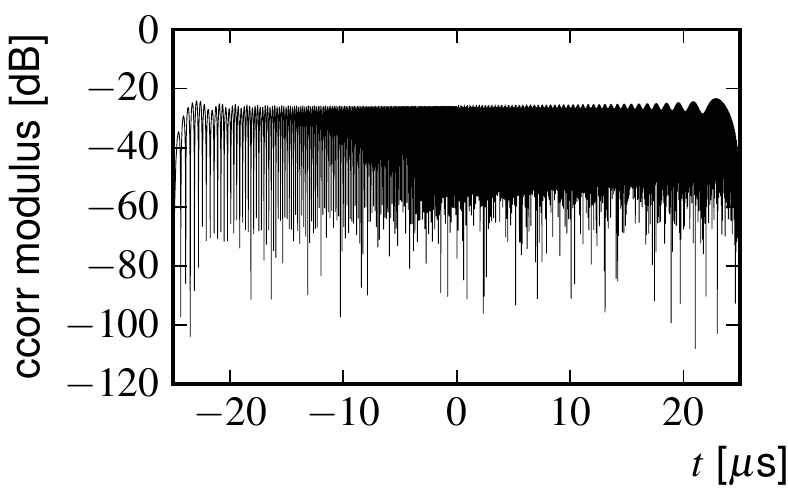}%
      }%
    }\hfill
    \subfloat[\label{sfig:randomFM}]{%
      \shortstack{%
        \includegraphics[scale=0.55]{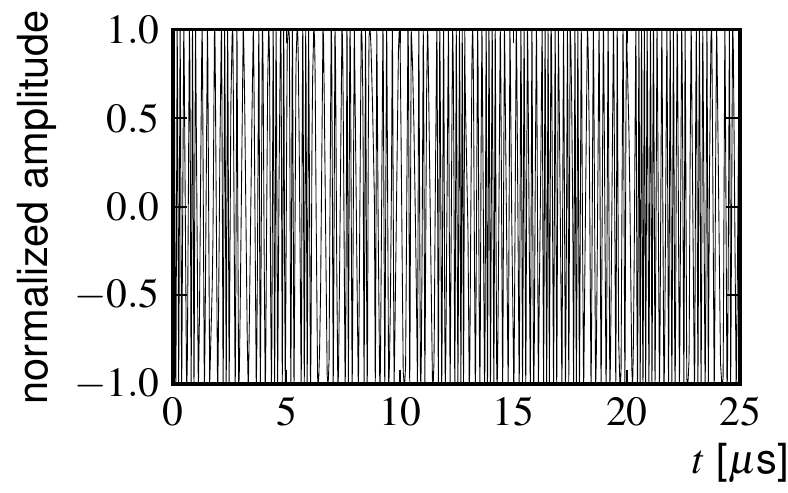}\\
        \includegraphics[scale=0.55]{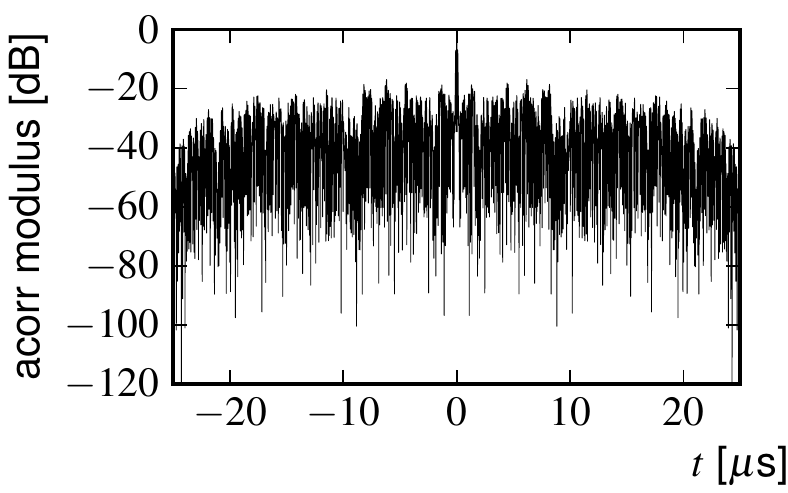}\\
        \includegraphics[scale=0.55]{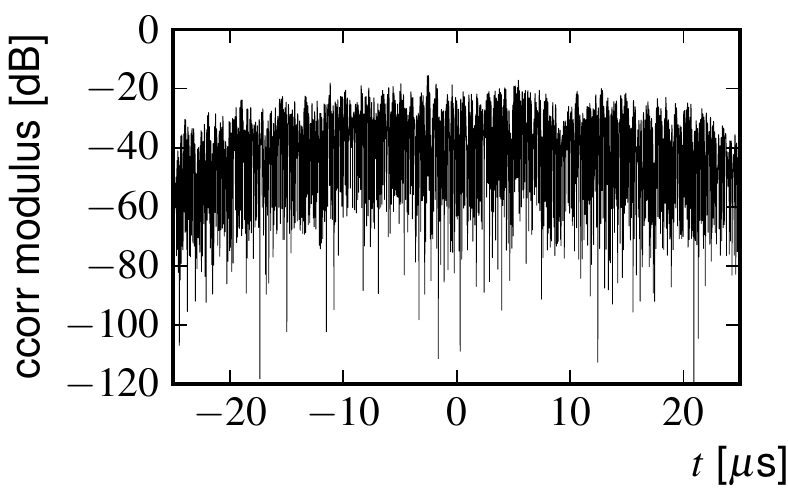}%
      }%
    }
  \end{center}
  \caption{From top to bottom: signal sample, auto-correlation and
    cross-correlation for a chirp \protect\subref{sfig:chirp} and a sample
    random-FM signal \protect\subref{sfig:randomFM}. Excitation lengths are
    artificially limited in the figure for representation purposes.}
  \label{fig:comparison}
\end{figure}

Fig.~\ref{sfig:chirp} shows a linear chirp (top), together with the modulus of
its acyclic auto-correlation (middle), that illustrates the suitability for
matched filter schemes, having a well localized central pulse and being rapidly
vanishes elsewhere.  Clearly, with this excitation it is impossible to have
arbitrary many transmitters simultaneously operated. The only family of
quasi-orthogonal functions that can be derived from it is made just of the
chirp and its reversed waveform. The modulus of their cross-correlation
(accounting for the mutual interference in a matched filter based receiver) is
shown at the bottom of Fig.~\ref{sfig:chirp}.

Random-FM excitations can be generated with the same modulator as chirps, given
in Eqn.~\eqref{eq:FM}, yet using a more articulated \ac{PAM} modulating
waveform
\begin{equation}
  x(t)= \sum_{k=-\infty}^{+\infty} x_k g(t - k\Tc)
  \label{eq:PAM}
\end{equation}
where $g(t)$ is a unit pulse of duration $\Tc$ and the values $x_k \in
[-1,+1]$ make up a random modulating sequence.

Fig.~\ref{sfig:randomFM} makes the parallel of Fig.~\ref{sfig:chirp} for a
sample random-FM excitation. Any detailed comparison must be postponed until
all the parameters influencing the modulation have been introduced (also due to
the rather artificial values used for representation).  Nonetheless, it is
already possible to see that arbitrary couples of random-FM excitations have
relatively low cross-correlations, since the \ac{PAM} sequence acts as a
\emph{signature} differentiating them. This makes random-FM suited for
\ac{MIMO}.

A thorough theory of random-FM signals has been developed in
\cite{Callegari:TCAS1-50-1}. Particularly, it is proved that the \ac{PDS} of an
(infinitely long) modulated signal depends on the \ac{PDF} of the modulating
sequence according to
\begin{multline}
  \label{eq:random-fm}
  \spectrum{s}(f) = 
  \int_{-1}^1 K_1(x,f-f_0)\rho(x)\,dx +\\
  \real\left(
    \frac{%
      \left(
        \int_{-1}^1 K_2(x,f-f_0)\rho(x)\, dx
      \right)^2}
    {1-\int_{-1}^1 K_3(x,f-f_0)\rho(x)\, dx}
  \right)
\end{multline}
where $\spectrum{s}(f)$ is the \ac{PDS}, the modulating sequence is assumed to
be made of independent samples, and $\rho(x)$ is its \ac{PDF}. Kernels
$K_1(x,f)$, $K_2(x,f)$, $K_3(x,f)$ are defined as
\begin{equation}
  \label{eq:kernels}
  \begin{split}
    K_1(x,f)&=\frac{1}{2}\Tc \sinc^2(\pi\Tc(f-\Df\,x))\\
    K_2(x,f)&=\ii 
    \frac{\ee^{-\ii 2 \pi \Tc (f-\Df x)}-1}{2\pi \sqrt{\Tc} (f-\Df x)}\\
    K_3(x,f)&=\ee^{-\ii 2 \pi \Tc (f-\Df x)}
  \end{split}
\end{equation}
Complicated as it may seem, the relationship can be interpreted as a
(nonlinear) \emph{smoothing and leaking} operator so that the $\spectrum{s}(f)$
tends to be shaped as $\rho(x)$ with some distortion.  For certain parameter
values, the relationship can be greatly simplified since the following
asymptotic tie holds:
\begin{equation}
  \label{eq:random-fm-slow}
  \lim_{\Tc\rightarrow\infty} \spectrum{s}(f) =
  \frac{1}{2\Df}\,\rho\left(\frac{f-f_0}{\Df}\right)
\end{equation}
Such tie, where the \ac{PDS} accurately \emph{copies} the \ac{PDF}, is also
\emph{approximately} valid for finite but large $\Tc$, namely when the \ac{PAM}
signal has a slow update frequency $f_c=\nicefrac{1}{\Tc}$. This case will be
indicated as a \emph{slow} modulation. Whether $\Tc$ makes the modulation
\emph{slow} or \emph{fast} depends on its magnitude relative to
$\nicefrac{1}{\Df}$. Thus, it is convenient to define a \emph{modulation index}
$m=\Tc\Df$. As long as $m$ is a few units or more, \eqref{eq:random-fm-slow}
provides a satisfactory approximation. Interestingly, using the tie in
Eqn.~\eqref{eq:random-fm} one can \emph{design} a modulating sequence \ac{PDF}
capable of producing a desired \ac{PDS}. The inversion can be practiced
directly through \eqref{eq:random-fm-slow} for the slow case or by iterative
methods (and reduced accuracy) for the fast case \cite{Callegari:TCAS1-50-8}.

\subsection{Applicability of random-FM signals to ultrasound \ac{NDT}}
Considering the applicability of random-FM signal to pulse compression, a few
items deserve attention.

\begin{asparaitem}
\item \emph{Auto-correlation away from the peak.} These correlation entries
  contribute to the noise floor at the receiving probes as self-interference,
  which may hide the detection of secondary echos.  From
  Fig.~\ref{fig:comparison}, it is evident that the auto-correlation is higher
  for the random-FM excitation than for the chirp, which may appear
  problematic. However, a few aspects are worth considering. First, the
  auto-correlation of the chirp keeps decreasing at a significant rate even at
  large lags. Conversely, the random-FM one only decreases rapidly for small
  lags, then it flattens. Thus, the best auto-correlation entries (large lags)
  are certainly much better for the chirp, yet this does not necessarily mean
  that the random-FM excitation is much worse in the worst case (small
  lags). Secondly, the higher auto-correlation profile is the price that one
  pays for the \ac{MIMO} abilities. Indeed, other coded excitations suitable
  for \ac{MIMO} also pay a price in auto-correlation with respect to the chirp
  \cite{Burrascano:CSIE-2009}. Furthermore, in a \ac{MIMO} setup, there is not
  just self-interference, but also mutual-interference that can be visualized
  through the cross-correlation entries. Having a self-interference much lower
  than the mutual-interference would only bring marginal advantages since the
  latter would dominate. Consequently, a fair evaluation of the random-FM
  auto-correlation curve also requires a comparison to the cross-correlation
  curves. Since the auto-correlation floor of the random-FM excitation is no
  worse than its cross-correlation floor or the cross-correlation floor of the
  chirp, one may conclude that the auto-correlation floor is actually
  \emph{good enough} since any improvement would only bring marginal advantages
  in a \ac{MIMO} setup.

\item \emph{Correlation dependency on system parameters}. Even if a complete
  discussion of this aspect is out of the scope of this paper, it is worth
  noticing that (as expectable) the correlation floors of random-FM modulations
  scale (in amplitude) with $\nicefrac{1}{\sqrt{\TB}}$ where $\TB$ is the
  time-bandwidth product of the excitation, approximately $2 T_e
  \Df$. Incidentally, the same relationship holds for the cross-correlation
  floor of \ac{PN} sequences such as Gold or Kasami. This provides some ability
  to tune the noise levels due to self- and mutual- interference to the
  application needs.

\item \emph{Secondary peaks in the correlation curves}. In
  Fig.~\ref{fig:comparison}, the correlation plots of the random-FM excitation
  appear irregular and peaky compared to those of the chirp. Secondary peaks
  are dangerous as they might be misinterpreted for weak echos from other
  reflectors at the receiving probes. Focusing on the auto-correlation curve,
  an intuitive investigation of their origin is possible. Auto-correlation is
  tied to \ac{PDS} by the Fourier transform. The \ac{PDS} of random-FM signals,
  as returned by Eqn.~\eqref{eq:random-fm} is, by the very properties of the
  involved operators, quite regular. The corresponding auto-correlation should
  necessarily be smooth. Thus, the peakiness in Fig.~\ref{sfig:randomFM}
  (middle) is not intrinsic in the random-FM waveforms. Its origin gets evident
  considering that Eqn.~\eqref{eq:random-fm} holds for infinitely long
  signals. Peakiness emerges as a short-length effect since the number of
  \ac{PAM} pulses used to build the excitation is limited, so that the value
  distribution of the modulating sequence can significantly differ from the
  prescribed \ac{PDF}. Similar considerations could hold for the peaks in the
  cross-correlations. Thus, the secondary peaks can be reduced by enlarging
  $\nicefrac{T_e}{\Tc}$. Being $\Tc=\nicefrac{m}{\Df}$, peaks can also be seen
  as a consequence of low $\nicefrac{(T_e\Df)}{m}$. Therefore, large $m$
  values, that from Section~\ref{ssec:background} may appear convenient for
  spectral control, can actually be undesirable. Note that in
  Fig.~\ref{sfig:randomFM} the peakiness is accentuated by the choice of very
  short excitations (\unit[25]{µs}) and relatively large $m$ (1), used for
  representation purposes.
\end{asparaitem}

\subsection{Specific advantages of random-FM excitations}
Established the applicability of random-FM excitations to \ac{NDT} (together
with the main limits in the choice of parameters), the specific pros of
random-FM excitation can finally be considered. Obviously, there is a first
advantage in using the same modulation type as chirps, since the implementation
can thus be based on quite similar hardware. Yet, the major asset of random-FM
modulations comes from the fine spectral shaping that can be practiced through
Eqns.~\eqref{eq:random-fm} and~\eqref{eq:random-fm-slow}.

\begin{figure}
  \begin{center}
    \subfloat[\label{sfig:proberesponse}]{%
      \includegraphics[scale=0.55]{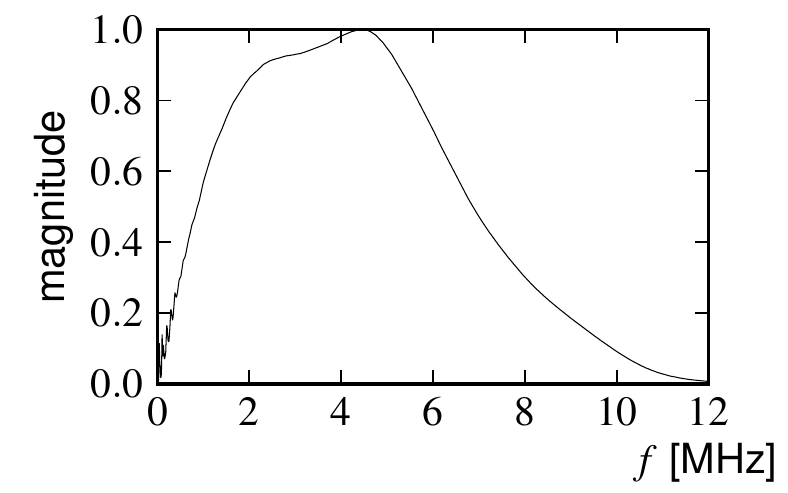}}
    \hfill
    \subfloat[\label{sfig:uniform-pds}]{%
      \includegraphics[scale=0.55]{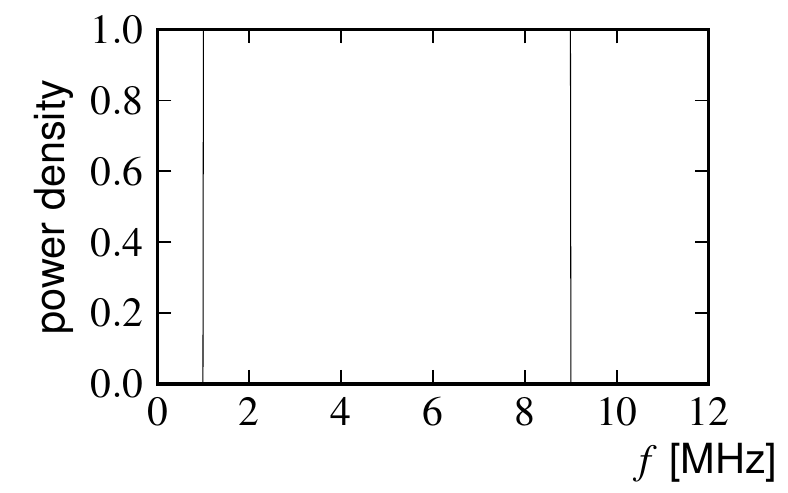}}
    \\
    \subfloat[\label{sfig:seconding-pds}]{%
      \includegraphics[scale=0.55]{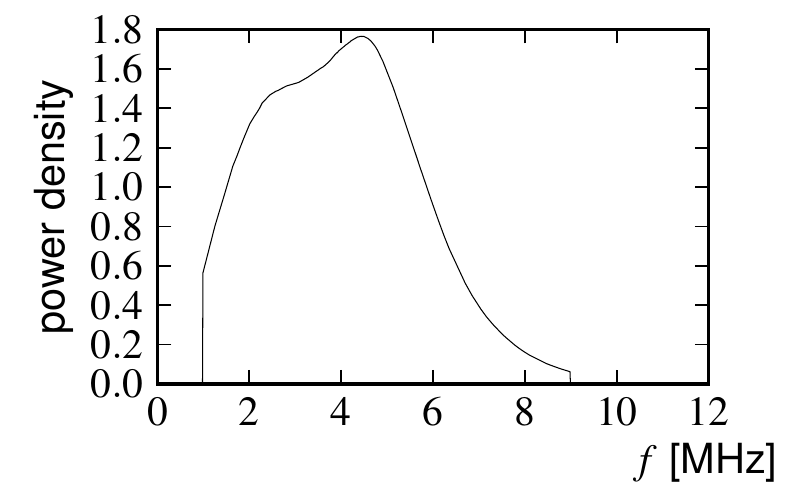}}
    \hfill
    \subfloat[\label{sfig:equalizing-pds}]{%
      \includegraphics[scale=0.55]{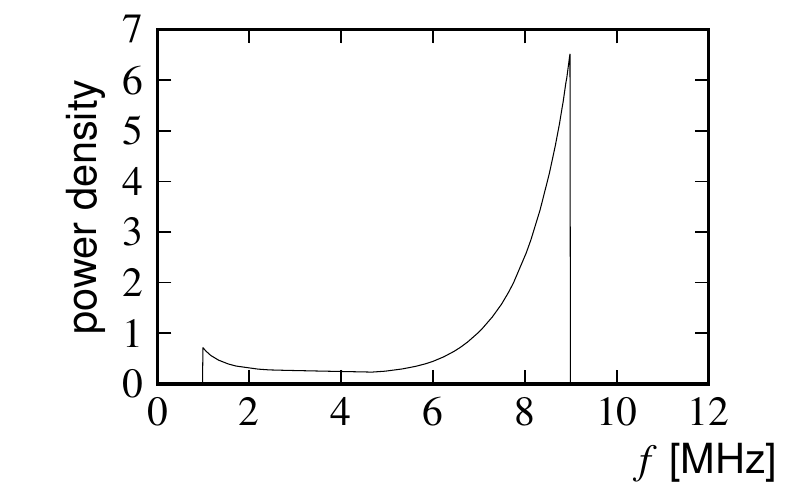}}
  \end{center}
  \caption{Overall normalized magnitude response of the probe system
    \protect\subref{sfig:proberesponse} and sample excitations power
    distributions: uniform \protect\subref{sfig:uniform-pds}; seconding the
    probe response \protect\subref{sfig:seconding-pds}; and equalizing the
    probe response \protect\subref{sfig:equalizing-pds}. Data on ordinate
    $\times 10^{-7}$.} 
\end{figure}

To appreciate this point, one should consider that the excitations are passed
to/from the \ac{MUT} through probes and amplifiers that are typically
characterized by non-flat frequency responses. For instance,
Fig.~\ref{sfig:proberesponse} illustrates the overall amplitude response of a
\unit[5]{MHz} probe system, as measured on field. If an excitation with a
\ac{PDS} uniform in the 1 to \unit[9]{MHz} range is adopted (shown
in~\ref{sfig:uniform-pds}), about 20\% of the excitation bandwidth and
about 43\% of the excitation power is lost in the probes.

Depending on the particular application setup, one may want a different
trade-off between the lost power and the lost bandwidth:

\begin{asparaenum}
\item If the \ac{eSNR} is critical and the noise is dominated by components
  that do not scale with the excitation power (i.e., components that are not
  self/mutual interference due to the transmitters, nor dispersion noise due to
  the intrinsic structure of the material, rather effects like thermal or
  quantization noise), reducing the power loss on the probes may be
  beneficial. Having the possibility to shape the excitation \ac{PDS}, this
  result can be achieved by picking a \ac{PDS} seconding the probe response, as
  shown in Fig.~\ref{sfig:seconding-pds}. This choice reduces the power loss to
  just 23\% with a minimal reduction in bandwidth.

\item Conversely, if the \ac{eSNR} is non critical, one may want to maximize
  the effective bandwidth. This can be done by picking a \ac{PDS} equalizing
  the probe response, as shown in Fig.~\ref{sfig:equalizing-pds}. This choice
  brings the actual excitation bandwidth up to the original \unit[8]{MHz}, but
  augments the power loss to 77\%.
\end{asparaenum}

It is worth recalling that a spectral shaping such as that in
plots.~\subref{sfig:seconding-pds} and~\subref{sfig:equalizing-pds} cannot be
practiced starting with a uniform-\ac{PDS} excitation and applying a linear
filter, because the filter would make the envelope of the excitation
non-constant. Conversely, random-FM signals can obtain the desired power
distribution quite simply by picking a modulating \ac{PAM} sequence with a
\ac{PDF} shaped as the desired spectrum. Indeed, being the required spectrum
(be it seconding or equalizing the probe response) typically smooth, either
slow or fast modulations can generally achieve it reasonably
well. Incidentally, note that in a recent past, a similar type of spectrum
shaping has been attempted (following the scenario at point 1), with non-linear
chirps \cite{Pollakowski:TUFFC-41-5}.

\begin{figure}
  \begin{center}
    \subfloat[\label{sfig:probabilities}]{%
      \includegraphics[scale=0.55]{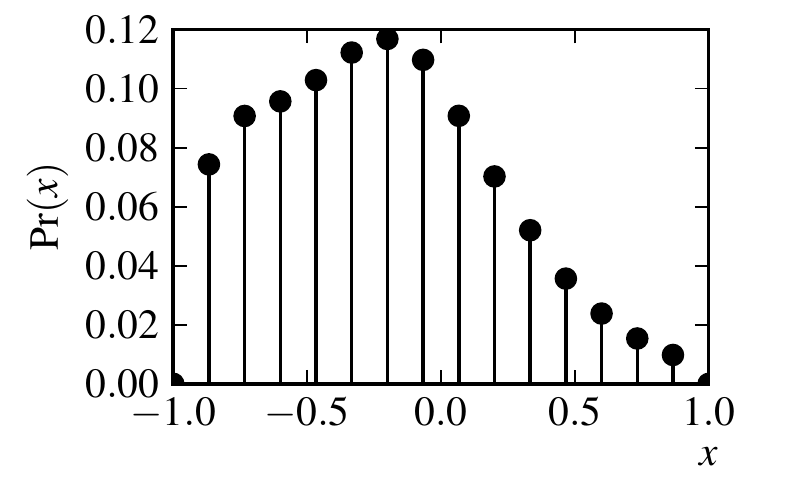}}%
    \hfill
    \subfloat[\label{sfig:achieved-pds}]{%
      \includegraphics[scale=0.55]{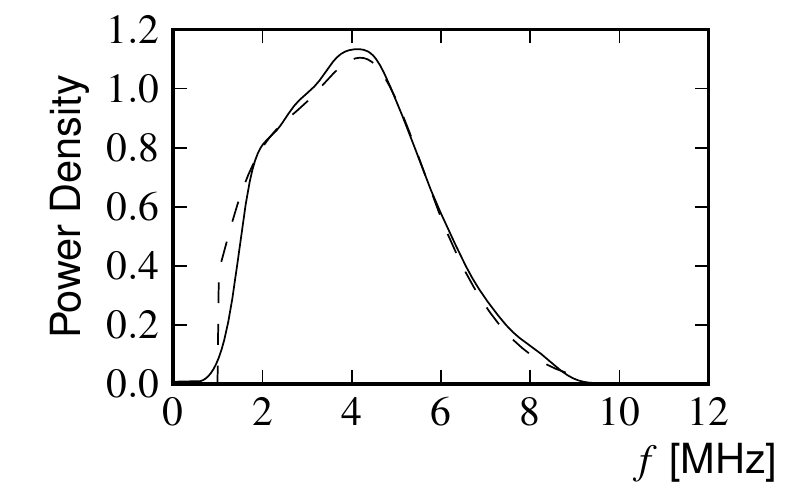}}
  \end{center}
  \caption{Use of a discrete-level \ac{PAM} modulating signal to achieve a
    desired \ac{PDS} in the random-FM waveform. In
    \protect\subref{sfig:probabilities} the probability distribution of a
    16-level \ac{PAM} signal. In \protect\subref{sfig:achieved-pds} the
    achieved \ac{PDS} at $m=4$ (solid line), together with the desired one
    (dashed), identical to that in Fig.~\ref{sfig:seconding-pds}.}
  \label{fig:discrete-PAM}
\end{figure}

As a last consideration, note that for the case of sufficiently fast
modulations, the smoothing properties of the operator in
Eqn.~\eqref{eq:random-fm} let one approximate reasonably well the desired
spectra using discrete-valued \ac{PAM} sequences. As an example,
Fig.~\ref{fig:discrete-PAM} shows the probability distribution required for a
16-levels \ac{PAM} modulation in order to obtain, at $m=4$, a random-FM signal
with a \ac{PDS} seconding the probe response. Pane~\subref{sfig:probabilities}
shows the probabilities, while pane~\subref{sfig:achieved-pds} shows the
achieved \ac{PDS} together with the desired one.

\section{Experimental evaluation and conclusions}
\begin{figure}
  \begin{center}
    \includegraphics[width=0.9\lw]{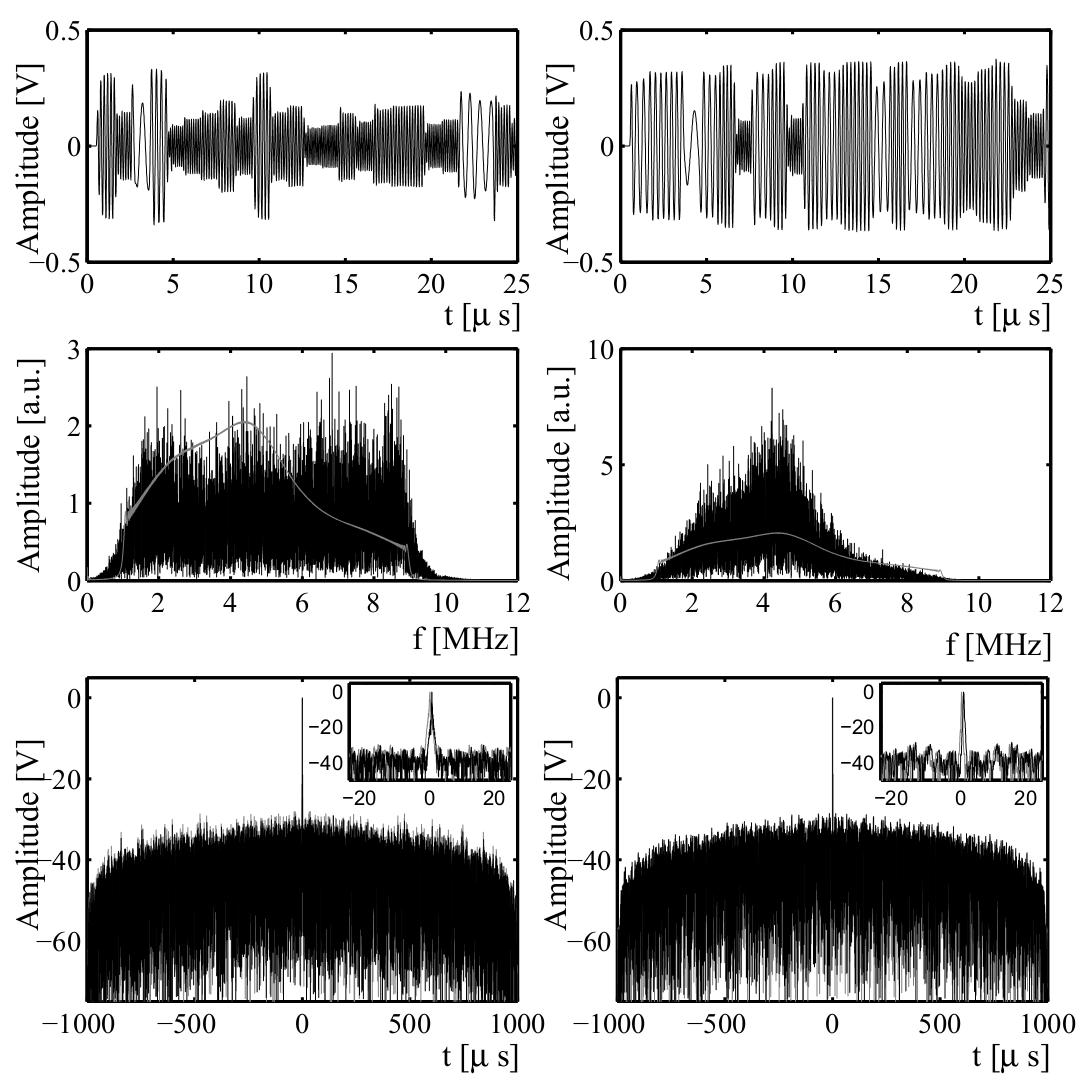}
  \end{center}
  \caption{Experimental results. Left: random-FM excitations designed for probe
    equalization; right: random-FM excitations designed for \ac{eSNR}
    enhancement. From top to bottom: measured signals at the receiving probes;
    corresponding spectra; auto-correlation. In the middle plots, the spectrum
    profiles that would be obtained by a linear chirp are superimposed in gray
    for reference, i.e., the characteristic spectrum of the two-probe
    system. In the bottom plots, expected trends are superimposed in gray.}
  \label{fig:experimental}
\end{figure}

The correlation properties and spectrum-shaping capability of random-FM
excitation have been experimentally validated by using a measurement set-up
composed by:
\begin{inparaenum}[(i)]
\item two broadband ultrasonic probes Olympus V109-RB centered at
  \unit[5]{MHz};
\item a National Instrument PXI-5412 Arbitrary Waveform Generator with
  generation rate up to \unitfrac[100]{Msample}{s} and \unit[14]{bit} of
  resolution;
\item a \unit[14]{bit} National Instrument PXIe-5122 Analog-to-Digital
  Converter with maximum sampling rate of \unitfrac[100]{Msample}{s}.
\end{inparaenum}
The two probes, one used as transmitter, the other as receiver, have been
arranged in through-transmission on a thin aluminum plate in order to
characterized the overall transmission channel. The transmitting probe has been
excited with a random-FM signal in compliance with Eqn.~\eqref{eq:FM}
and~\eqref{eq:PAM} and characterized by the following parameters
$T_e$=\unit{1}{ms}; $f_0$=\unit[5]{MHz}; $\Df$=\unit[4]{MHz}; amplitude
$A$=\unit[6]{V}; $m$=8. Signals have been digitally generated and acquired at a
sampling rate of \unitfrac[100]{Msample}{s}. Experimental data is summarized in
Fig.~\ref{fig:experimental}.

Acquired signals at the receiving probes (top), corresponding spectra (middle)
and retrieved auto-correlation functions (bottom) are illustrated for the two
cases of random-FM modulation finalized at equalizing the probe response or
seconding it.  The correspondence between numerical simulations and
experimental data is quite good and supports the possibility of exploiting
random-FM signals in \ac{NDT} ultrasonic \ac{MIMO} systems. Specifically, the
ability of the excitation seconding the probe response to enhance the received
power is quite visible. Conversely, at least in the tested setup, the
equalization of the probe response does not appear particularly advantageous.
The features of the auto- and cross- correlation functions, with specific
regard to the noise floor level, make this approach suitable in applications
where single echoes have to be detected by each transmitter-receiver pair, such
as ultrasonic localization, time of flight measurement, etc., and several pairs
are needed. On the other hand, in those applications that require a high ratio
between the main lobe of the auto-correlation and the noise floor, such as
certain defect-detection systems, chirp signals may still be preferable.
Strategies to reduce the noise floor are currently under investigation,
involving a deep analysis of the relations between modulating function
features, time-frequency behavior of signals and noise floor levels.

\section*{Acknowledgment}
{\small Work funded by MIUR PRIN 2009 project ``Diagnostica non distruttiva ad
  ultrasuoni tramite sequenze pseudo-ortogonali per imaging e classificazione
  automatica di prodotti industriali'' (USUONI). M.R. acknoledges partial
  financial support from Fondazione CARIT.}




%

\bibliographystyle{IEEEtran}
\bibliography{macros,IEEEabrv,ultrasound,various,sensors,analog,chaos}

\begin{thebibliography}{1}
\providecommand{\url}[1]{#1}
\csname url@samestyle\endcsname
\providecommand{\newblock}{\relax}
\providecommand{\bibinfo}[2]{#2}
\providecommand{\BIBentrySTDinterwordspacing}{\spaceskip=0pt\relax}
\providecommand{\BIBentryALTinterwordstretchfactor}{4}
\providecommand{\BIBentryALTinterwordspacing}{\spaceskip=\fontdimen2\font plus
\BIBentryALTinterwordstretchfactor\fontdimen3\font minus
  \fontdimen4\font\relax}
\providecommand{\BIBforeignlanguage}[2]{{%
\expandafter\ifx\csname l@#1\endcsname\relax
\typeout{** WARNING: IEEEtran.bst: No hyphenation pattern has been}%
\typeout{** loaded for the language `#1'. Using the pattern for}%
\typeout{** the default language instead.}%
\else
\language=\csname l@#1\endcsname
\fi
#2}}
\providecommand{\BIBdecl}{\relax}
\BIBdecl

\bibitem{Gan:Ultrasonics-39-3}
T.~H. Gan, D.~A. Hutchins, D.~R. Billson, and S.~D. W., ``The use of broadband
  acoustic transducers and pulse-compression techniques for air-coupled
  ultrasonic imaging,'' \emph{Ultrasonics}, vol.~39, no.~3, pp. 181 -- 194,
  2001.

\bibitem{Turin:IRETIT-6-3}
G.~Turin, ``An introduction to matched filters,'' \emph{IRE Transactions on
  Information Theory}, vol.~6, no.~3, pp. 311 -- 329, Jun. 1960.

\bibitem{Callegari:CAT-2005-5-alt}
S.~Callegari, G.~Mazzini, R.~Rovatti, and G.~Setti, ``A chaos approach to
  asynchronous {DS}-{CDMA} systems,'' in \emph{Chaos Applications in
  Telecommunications}, P.~Stavroulakis, Ed.\hskip 1em plus 0.5em minus
  0.4em\relax Boca Raton, FL, USA: CRC International Press, 2005, ch.~5, pp.
  187--221.

\bibitem{Golomb:SDGC-2005}
S.~W. Golomb and G.~Gong, \emph{Signal Design for Good Correlation: For
  Wireless Communication,Cryptography,and Radar}.\hskip 1em plus 0.5em minus
  0.4em\relax Cambridge University Press, 2005.

\bibitem{Burrascano:CSIE-2009}
P.~Burrascano, A.~Pirani, and M.~Ricci, ``Exploiting pseudo orthogonal
  pn-sequences for ultrasonic imaging system,'' in \emph{Proceedings of 2009
  WRI World Congress on Computer Science and Information Engineering}, 2009,
  pp. 181 -- 185.

\bibitem{Callegari:TCAS1-50-1}
S.~Callegari, R.~Rovatti, and G.~Setti, ``Spectral properties of chaos-based
  {FM} signals: Theory and simulation results,'' \emph{{IEEE} Trans. Circuits
  Syst. {I}}, vol.~50, no.~1, pp. 3--15, Jan. 2003.

\bibitem{Ricci:IUS-2012}
M.~Ricci, S.~Callegari, S.~Caporale, M.~Monticelli, M.~Eroli, L.~Senni,
  R.~Rovatti, G.~Setti, and P.~Burrascano, ``Exploiting non-linear chirp and
  sparse deconvolution to enhance the performance of pulse-compression
  ultrasonic {NDT},'' in \emph{Proceedings of IUS}, Dresden (DE), Oct. 2012.

\bibitem{Pollakowski:TUFFC-41-5}
M.~Pollakowski and H.~Ermert, ``Chirp signal matching and signal power
  optimization in pulse-echo mode ultrasonic nondestructive testing,''
  \emph{{IEEE} Trans. Ultrason., Ferroelectr., Freq. Control}, vol.~41, no.~5,
  pp. 655 -- 659, Sep. 1994.

\bibitem{Callegari:TCAS1-50-8}
S.~Callegari, R.~Rovatti, and G.~Setti, ``Chaos based {FM} signals:
  Applications and implementation issues,'' \emph{{IEEE} Trans. Circuits Syst.
  {I}}, vol.~50, no.~8, pp. 1141--1147, Aug. 2003.

\end{thebibliography}
\end{document}